\documentclass{ws-p8-50x6-00}

\begin{document}
%
%
\def\koko{\mbox{K}_s^0\mbox{K}_s^0 }
\def\ra{\rightarrow }
\def\gam{\gamma }
\def\Ggg{\Gamma_{\gamma\gamma} }
\def\epem{\mbox{e}^+\mbox{e}^- }
\def\pip{\pi^+ }
\def\pim{\pi^- }
\def\ko{\mbox{K}^0 }
\def\kobar{\bar{\mbox{K}^0} }
\def\kol{\mbox{K}^0_L }
\def\kos{\mbox{K}^0_S }
\def\k{\mbox{K}}      
\def\pb{pb$^{-1}$}
\def\NP{{\it Nucl. Phys. }}
\def\PL{{\it Phys. Lett. }}
\def\ZfP{{\it Z. Phys. }}
\def\NIM{{\it Nucl. Inst. Meth. }}
\def\PRep{{\it Phys. Rep. }}
\def\PR{{\it Phys. Rev. }}
\def\PRL{{\it Phys. Rev. Lett. }}
\def\EURO{{\it Eur. Phys. J. }}

\title{
Resonances and Exclusive Channels:\\
an Experimenter's Summary
}

\author{S. Braccini
\footnote{Talk given at Photon 2001, Ascona, Switzerland, September 2001.} 
}

\address{INFN, Laboratori Nazionali di Frascati, \\
Via E. Fermi, 40 - I-00044 Frascati (Rome), Italy\\
E-mail: Saverio.Braccini@cern.ch}


\maketitle

\abstracts{
 A very remarkable number of new results in the study of resonances and 
exclusive channels has been presented at this conference giving
fundamental information in the understanding of strong interactions at low
energies. The first results
from the new high luminosity colliders are impressive and a lot of
activity in this field is foreseen for the future. 
The most relevant issues are summarized and discussed in this paper. 
}

\section{Introduction}

 The study of resonances and exclusive channels mostly aims to understand
the behaviour of the strong interactions in the energy range up to a few GeV.
Here perturbative QCD cannot in general be applied and phenomenological models are 
needed to interpret the data which are often used as input to the models themselves.
In this energy range the quarks and the antiquarks form
a large number of bound states which are commonly interpreted as mesons and baryons.

 In this scheme a fundamental question is still open: do glueballs exist?
According to QCD, bound states of one or more gluons can be formed but a solid
experimental observation is still missing. The search for
this form of matter made only by boson force carriers is one of the most actual
themes of research in this field.

  Because of the large mass of the charm quark, the study 
of the formation of charmonium states allows to test
non-relativistic perturbative QCD calculations and to measure $\alpha_s$ at the charm scale.

 Two-photon collisions at electron positron storage rings represent
a very good and clean environment for this kind of studies~\cite{SaverioMeson2000}. 
Since gluons do not couple directly to photons, the two-photon process is a powerfull
glueball anti-filter.

 In the last few years a very remarkable progress has been achieved due to the results
of many experiments. Considering two-photon physics, LEP at CERN and CESR at Cornell have
produced a large number of important results. At this conference the first results from the
new e$^+$e$^-$ high luminosity machines DA$\Phi$NE at LNF and BELLE at KEK have been
presented. These results are impressive and sometimes already comparable
with the achievements of the machines of the previous generation. 
A bright future for this field of research is therefore foreseen.

\section{Lepton, meson and baryion pair production in two-photon collisions}

 The study of lepton pair production is a test for QED
and an important tool to understand experimental apparates. The production of muon~\cite{Debreczeni}
and $\tau$ pairs~\cite{Haas} is studied by L3 and bounds for anomalous couplings of the $\tau$ lepton
are set for the first time in this kind of studies.

 Charged kaon and pion pair production in two-photon collisions is studied by ALEPH 
and DELPHI~\cite{Grzelak}. Good agreement is found between the experimental results and the
Brodsky-Lepage model for the kaons. For the pions there is not good agreement either between 
the two experimental measurements or between experiment and theory. 
More work is therefore required to understand the data. 

 The study of baryon-antibaryon pairs allows to test the predictions of pure quark
and quark-diquark models. From the study of the reaction $\gamma\gamma\rightarrow$p$\bar{\rm p}$ 
presented by OPAL~\cite{Barillari} it is difficult to have an indication on 
which model reproduces better the data. 
The preliminary results of the same channel presented by BELLE~\cite{Chen} show that this channel 
will be studied with a much larger data sample in the near future. 
The production of $\Lambda$ and $\Sigma^0$ baryon pairs is studied by L3~\cite{Echenard} and 
good agreement is found with the quark-diquark model predictions. The pure quark model is
disfavoured by this analysis. 
 
\section{Pseudoscalar and vector mesons}

 The precise measurements of BR($\Phi\rightarrow\eta'\gamma$) and of the ratio
BR($\Phi\rightarrow\eta'\gamma$)/BR($\Phi\rightarrow\eta\gamma$) by KLOE~\cite{Lanfranchi}
disfavor models with a large gluonium content in the $\eta'$. This is in agreement with
the previous results obtained by L3~\cite{L3-etap} in two-photon collisions.

 The $\kos{\rm K}^\pm\pi^\mp$ and $\eta\pi^+\pi^-$ final states in two-photon collisions are
studied by L3~\cite{Vodopianov}. The formation of the $\eta(1440)$ and of the
f$_1$(1420) as a function
of $Q^2$ is investigated for the first time, showing that also vector mesons can be studied
using the two-photon process if a large data sample is available.
The two-photon width of the $\eta(1440)$ is obtained using data at low $Q^2$, as reported in
Table~\ref{summary}. 
This first observation of the $\eta(1440)$ in untagged two-photon
collisions disfavors its interpretation as the 0$^{-+}$ glueball in agreement with the lattice QCD
calculations. The $\eta(1440)$ can therefore be interpreted as a radial excitation~\cite{Anisovitch-1}.

 The first sign of the production of an $\eta_b$ meson may have been shown for the
first time at this conference by ALEPH~\cite{Boherer}.

\section{Scalars, tensors and glueball searches}

\begin{table}[b]
\caption{Spin and helicity studies of the ${\rm K}\bar{\rm K}$ final state by L3 and BELLE.}
\begin{tabular}{|l|c|c|c|}
\hline
    & $\kos\kos$ (L3) & $\kos\kos$ (BELLE)  & ${\rm K}^+{\rm K}^-$ (BELLE)\\
\hline 
f$_2'(1525)$   & (J=2, $\lambda$=2) only & (J=2, $\lambda$=2) only & (J=2, $\lambda$=2) only\\
1750 MeV       & (J=2, $\lambda$=2) dominant	& (J=0) dominant & (J=2, $\lambda$=0) dominant\\
\hline
\end{tabular}
\label{spin-helicity}
\end{table}

 The tensor meson nonet is well established and is nowadays used as a test
for other measurements. 
On the other hand, the interpretation of the scalar meson
nonet is still an open and important problem to be solved.
According to lattice QCD predictions~\cite{Bali}, the ground state glueball is
a scalar with a mass between 1400 and 1800 MeV and 
the tensor glueball is heavier with a mass between  1900 and 2300 MeV.
Since several 0$^{++}$ states have been observed in the 1400-1800 MeV
mass region, the scalar ground state glueball can  mix with nearby quarkonia,
making the search for the scalar glueball and the interpretation
of the scalar meson nonet a single problem~\cite{Amsler}.
The interpretation of the 1400-1800 MeV mass region is made even more difficult
by the fact that radially excited tensor meson states are also predicted in this
mass region.

 The $\pi^+\pi^-\pi^0$ final state in two-photon collisions, studied by L3~\cite{Levtchenko}
and BELLE~\cite{Hou}, is dominated by the formation of the a$_2$(1320). These studies
clearly confirm the observation of the radially excited tensor meson
a$_2'$(1750), already reported by L3 in a previous study~\cite{L3-a2}. The values obtained for
the two-photon width (Table~\ref{summary}) by the two experiments are consistent and agree with the 
theoretical predictions~\cite{Munz}. The spin-parity analysis performed by L3 shows some indications
for other states which will be possibly put in evidence in future as soon as larger data
samples will be available.

 The K$\bar{\rm K}$ final state in two-photon collisions is largely dominated by resonance formation 
and is therefore one of the golden channels in the study of scalar and tensor states. 
A final study of the $\kos\kos$ final state is
reported by L3~\cite{L3-k0k0} and a preliminary study of the $\kos\kos$ and the  K$^+$K$^-$ 
final states is presented by BELLE~\cite{Uehara}.
The two $\kos\kos$ mass spectra show nearly identical features. Around 1300 MeV a small signal is
due to the f$_2$(1270)-a$_2$(1320) destructive interference. The spectrum is dominated by the formation f$_2'$(1525)
tensor meson for which the two-photon width is measured with high precision by L3,
as reported in Table~\ref{summary}. A very clear signal is
present around 1750 MeV, exactly where the $s\bar{s}$ member of scalar meson nonet and the radially excited
tensor mesons are expected. To investigate the spin J and the helicity $\lambda$, 
the decay angular distributions are studied, as reported in Table~\ref{spin-helicity}. Good
agreement is found only in the f$_2'(1525)$ mass region while the interpretation of the 1750 MeV mass region
is still unclear. L3 reports a measurement of the two-photon width of the 
f$_2$(1750) (Table~\ref{summary}), in agreement with the theoretical predictions for the radially excited
tensor mesons~\cite{Munz}.
More than one wave is very probably present in this mass region. The presence of
a J=0 state reported by L3 with a fraction of 24$\pm$16\% and by BELLE, if confirmed, is very important
to support the interpretation of the f$_0$(1300), the f$_0$(1500) and f$_0$(1750) as the two isoscalar
members of the scalar nonet mixed with the ground state glueball~\cite{Amsler}. 
If this is the case, the f$_0$(980)
and the a$_0$(980) cannot be considered as $q\bar{q}$ states, 
as suggested by the recent results by KLOE~\cite{Lanfranchi}
in the study of the decays $\Phi\ra{\rm a}_0(980)\gamma$ and $\Phi\ra{\rm f}_0(980)\gamma$.

 No evidence for the observation of the narrow $\xi(2230)$ tensor glueball candidate~\cite{PDG}
is reported by L3~\cite{L3-k0k0} and BELLE~\cite{Uehara}. Upper limits
for the two-photon width of the $\xi(2230)$ are derived,
in agreement with the results by CLEO~\cite{CLEO-xi} (Table~\ref{summary}).
This is in favour of the interpretation of the $\xi(2230)$
as the tensor glueball in case of a confirmation in gluon rich environments or 
is just an indication that this state simply does not exist.

\section{Charmonia}

 New preliminary results on the
formation of the $\eta_c(2980)$ have been submitted by DELPHI~\cite{DELPHI-etac}.
The two-photon width is measured using the $\kos{\rm K}^\pm\pi^\mp$, 
${\rm K}^+{\rm K}^-{\rm K}^+{\rm K}^-$ and ${\rm K}^+{\rm K}^-\pi^+\pi^-$
decay modes leading to the combined result reported in Table~\ref{summary}.
This measurement is in good agreement with the
previous measurements~\cite{PDG} and with the theoretical 
predictions~\cite{Fabiano}. No signal of the $\eta_c$
is observed in the $\pi^+\pi^-\pi^+\pi^-$ final state and the upper limit 
$\Gamma_{\gamma\gamma}(\eta_c)<3.8\,\,{\rm keV}$ is derived. This contradictory
problem is still under investigation.

 Two new measurements of the two-photon width of the $\chi_{c2}$ have been presented 
in this conference by BELLE~\cite{Uehara} and CLEO~\cite{Paar}, as presented in Table~\ref{summary}.
It is interesting to remark that the measurement by BELLE is based on
the ``usual'' J/$\psi\gamma$ decay mode while 
CLEO performs the first measurement using the $\pi^+\pi^-\pi^+\pi^-$ decay mode.

 The measurements of the two-photon width of the $\eta_c$ performed in two-photon collisions
are in good agreement with the ones obtained in ${\rm p}\bar{{\rm p}}$ annihilations~\cite{PDG}.
The situation is different for the $\chi_{c2}$. 
The measurements of the two-photon width of the $\chi_{c2}$ performed in two-photon collisions
and using the  J/$\psi\gamma$ decay mode
are significantly higher than the values measured in ${\rm p}\bar{{\rm p}}$
annihilations~\cite{PDG}.
The reason for this could be due to a systematic effect affecting the two-photon measurements
based on the J/$\psi\gamma$ decay mode.
As a matter of fact, this new measurement by CLEO is surely not affected by the same systematic effects
and is in better agreement with the measurements obtained in ${\rm p}\bar{{\rm p}}$ annihilations.

 The first measurement of the two-photon width of the $\chi_{c0}$
is obtained by CLEO~\cite{Paar}. The result is reported in Table~\ref{summary} and
is obtained using the $\pi^+\pi^-\pi^+\pi^-$ decay mode.
An indication for the formation of the $\chi_{c0}$ is present in the $\kos\kos$
mass spectrum presented by BELLE~\cite{Uehara}. If confirmed, it will be interesting
to have the possibility to perform a completely independent measurement of the two-photon
width of the $\chi_{c0}$.

\begin{table}[t]
\caption{The most recent results on the two-photon width of mesons, charmonia, radial excitations and glueball
candidates. 
($\dag$ the value is given times the decay branching ratio)}
\begin{tabular}{|l|c|c|c|c|c|}
\hline
Resonance  & Experiment & Final state  & J$^{PC}$& $\Gamma_{\gamma\gamma}$ & Ref. \\
\hline 
 $\eta'$(958)         & L3             & $\pi^+ \pi^- \gamma$ & $ 0^{-+}   $ & 4.17$\pm$0.10$\pm$0.27 keV & ~\cite{L3-etap}\\
 a$_2$(1320)          & L3             & $\pi^+ \pi^- \pi^0 $ & $ 2^{++}   $ & 0.98$\pm$0.05$\pm$0.09 keV & ~\cite{L3-a2} \\
 f$_2^{'}$(1525)      & L3             & K$^0_s$K$^0_s$       & $ 2^{++}   $ & 0.085$\pm$0.007$\pm$0.012 keV & ~\cite{L3-k0k0}\\ \hline
 $\eta _{c}$(2980)    & L3	       & 9 chan.  	      & $ 0^{-+}   $ & 6.9$\pm$1.7.$\pm$0.8  keV & ~\cite{L3-etac}\\
 $\eta _{c}$(2980)    & DELPHI	       & 3 chan.  	      & $ 0^{-+}   $ & 13.0$\pm$2.7.$\pm$5.0  keV & ~\cite{DELPHI-etac}\\
 $\eta _{c}'$         & L3	       & 9 chan.  	      & $ 0^{-+}   $ & $<2.0$ keV  & ~\cite{L3-etac}\\
 $\chi_{c2}$(3555)    & L3	       & $ l^+ l^- \gamma $   & $ 2^{++}   $ & 1.02$\pm$0.40$\pm$0.15 keV & ~\cite{L3-chic}\\ 
 $\chi_{c2}$(3555)    & OPAL	       & $ l^+ l^- \gamma $   & $ 2^{++}   $ & 1.76$\pm$0.47$\pm$0.37 keV & ~\cite{OPAL-chic}\\ 
 $\chi_{c2}$(3555)    & BELLE	       & $ l^+ l^- \gamma $   & $ 2^{++}   $ & 0.84$\pm$0.08$\pm$0.10 keV & ~\cite{Uehara}\\ 
 $\chi_{c2}$(3555)    & CLEO	       &$\pi^+\pi^-\pi^+\pi^-$& $ 2^{++}   $ & 0.53$\pm$0.15$\pm$0.23 keV & ~\cite{Paar}\\ 
 $\chi_{c0}$(3415)    & CLEO	       &$\pi^+\pi^-\pi^+\pi^-$& $ 2^{++}   $ & 3.76$\pm$0.65$\pm$1.81 keV & ~\cite{Paar}\\ \hline
 $\eta$(1440)         & L3	       & K$^0_s$K$^\pm\pi^\mp$& $ 0^{-+}   $ & 0.199$^{\dag}\pm$0.52 keV& ~\cite{Vodopianov}\\
 f$_2$(1750)	      & L3	       & K$^0_s$K$^0_s$       & $ 2^{++}   $ & 0.049$^{\dag}\pm$0.011$\pm$0.013 keV & ~\cite{L3-k0k0}\\
 a$_2'$(1752)         & L3	       & $\pi^+ \pi^- \pi^0 $ & $ 2^{++}   $ & 0.29$^{\dag}\pm$0.04$\pm$0.02 keV & ~\cite{L3-a2}\\ 
 a$_2'$(1752)         & BELLE	       & $\pi^+ \pi^- \pi^0 $ & $ 2^{++}   $ & 0.27$^{\dag}\pm$0.02$\pm$0.04 keV & ~\cite{Hou}\\ \hline
 f$_0$(1500)	      & ALEPH	       & $\pi^+ \pi^-$        & $ 0^{++}   $ & $<310^{\dag}$ eV & ~\cite{ALEPH-pipi}\\
 f$_0$(1710)	      & ALEPH	       & $\pi^+ \pi^-$        & $ 0^{++}   $ & $<550^{\dag}$ eV & ~\cite{ALEPH-pipi}\\
 $\xi$(2230)	      & CLEO	       & $\pi^+ \pi^-$        & $ 2^{++}   $ & $<2.5^{\dag}$ eV& ~\cite{CLEO-xi}\\
 $\xi$(2230)	      & CLEO	       & K$^0_s$K$^0_s$       & $ 2^{++}   $ & $<1.3^{\dag}$ eV& ~\cite{CLEO-xi}\\
 $\xi$(2230)	      & L3	       & K$^0_s$K$^0_s$       & $ 2^{++}   $ & $<1.4^{\dag}$ eV& ~\cite{L3-k0k0}\\ \hline
\end{tabular}
\label{summary}
\end{table}

\section{Conclusions and outlook}

 The study of resonances and exclusive channels is a very interesting
and active field of research.
A remarkable progress on the study of resonance formation in
two-photon collisions has been achieved in the last few years.
Data from  the LEP collider at CERN and CESR at Cornell allowed to
improve significantly the precision on the two-photon widths of
several resonances, to identify
some radial excitations and to search for glueball candidates.
New high luminosity $\epem$ machines have been built and are now
starting their data taking. The first results from BELLE at KEK and 
KLOE at LNF represent a good sign for the future and new projects like CLEO-c~\cite{Cassel} are welcome.
The most relevant recent results on resonance formation in two-photon collisions
are summarized in Table~\ref{summary} and
represent an important contribution to meson
spectroscopy and glueball searches.

\section*{Acknowledgments}

 I would like to acknowledge all the organizers of Photon2001
for the very nice and friendly atmosphere we had in Ascona.
I would like to thank in particular G. Bali, co-convener of this session,
M.N. Focacci-Kienzle, J.H. Field, M. Wadhwa
for the very constructive discussions and suggestions.

\end{document}